\documentclass[conference]{IEEEtran}
\usepackage{cite}
\usepackage{graphicx}
\usepackage{enumerate}
\usepackage[english]{babel}
\usepackage[T1]{fontenc}
\usepackage{listings}
\usepackage{xcolor}

\ifCLASSINFOpdf
\else
\fi

\begin{document}
%
\title{Design and implementation of a digital clock showing digits in Bangla font using microcontroller AT89C4051 }

\author{\IEEEauthorblockN{Nasif Muslim, Md. Tanvir Adnan, Mohammad Zahidul Kabir, Md. Humayun Kabir, Sheikh Mominul Islam}
\IEEEauthorblockA{American International University-Bangladesh (AIUB),
 Bangladesh\\
					}
}


\maketitle

\begin{abstract}
In this paper, a digital clock is designed where the microcontroller is used for timing controller and the 
font of the Bangla digits are designed, and programmed within the microcontroller. The design is cost effective, 
simple and easy for maintenance.
\end{abstract}


%
\IEEEpeerreviewmaketitle

\section{Introduction}
A clock  is an instrument for measuring time. In principle, it requires no more than some physical process which will proceed at a known rate, and a way to gauge how long that process has been continuing. There are different types and sizes of clocks are available. \\

Depending on the method of time display, clocks can be classified by four types. They are i) analog clocks, ii) digital clocks, iii) auditory clocks, iv) textual clocks. Digital clocks display a numeric representation of time. Two numeric display formats are commonly used on digital clocks. They are: 24-hour notation with hours ranging 00 to 23 and  12 hour notation with AM/PM indicator. Most digital clocks use an LCD or LED display. \\

Generally for the designing a digital clock, a microcontroller is used as the controller of the circuit and a Real Time Clock IC \cite{dallas} is used as counter. But for displaying a Bangla digit character set in Seven Segment Display \cite{avago}, it is required to use about a hundred or more NAND/NOR gate. It is not feasible to use so many gates and it will increase the size of the hardware too. To solve this problem, bangla digits are designed and programmed in the microcontroller and LCD display \cite{lcd} is used as a display.

\section{Software Design}	
The software design process consists of two steps. 
	\begin{enumerate}
		\item Bangla digit design 
		\item Clock algorithm design
	\end{enumerate}

\subsection{Bangla digit design}
Bascom-8051 \cite{bascom} is used to design and simulate Bangla digits. It is a windows BASIC compiler for the AVR family microcontroller. The main advantage of using this software is that, it has an LCD designer by which it is possible to design custom characters or digits for LCD display. It is also possible to examine the custom design using LCD simulator. 

\begin{figure}[htbp]
	\begin{center}
	 \includegraphics[angle=0,scale=.45]{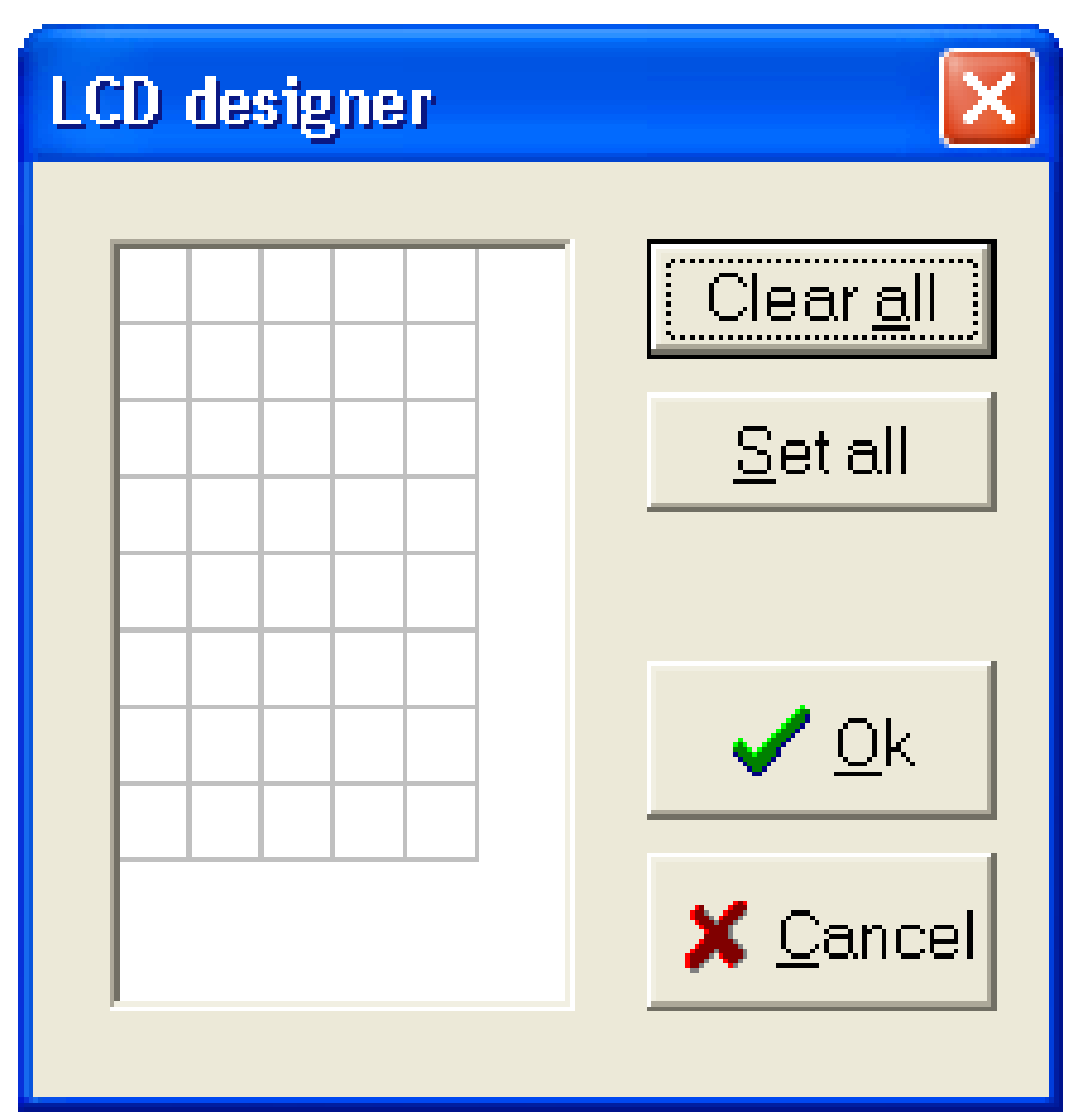}
	\end{center}
	\caption{ LCD designer - the LCD matrix has 7x5 cells  }
	\label{LCD designer1} 
\end{figure}

\begin{figure}[htbp]
	\begin{center}
	 \includegraphics[angle=0,scale=.58]{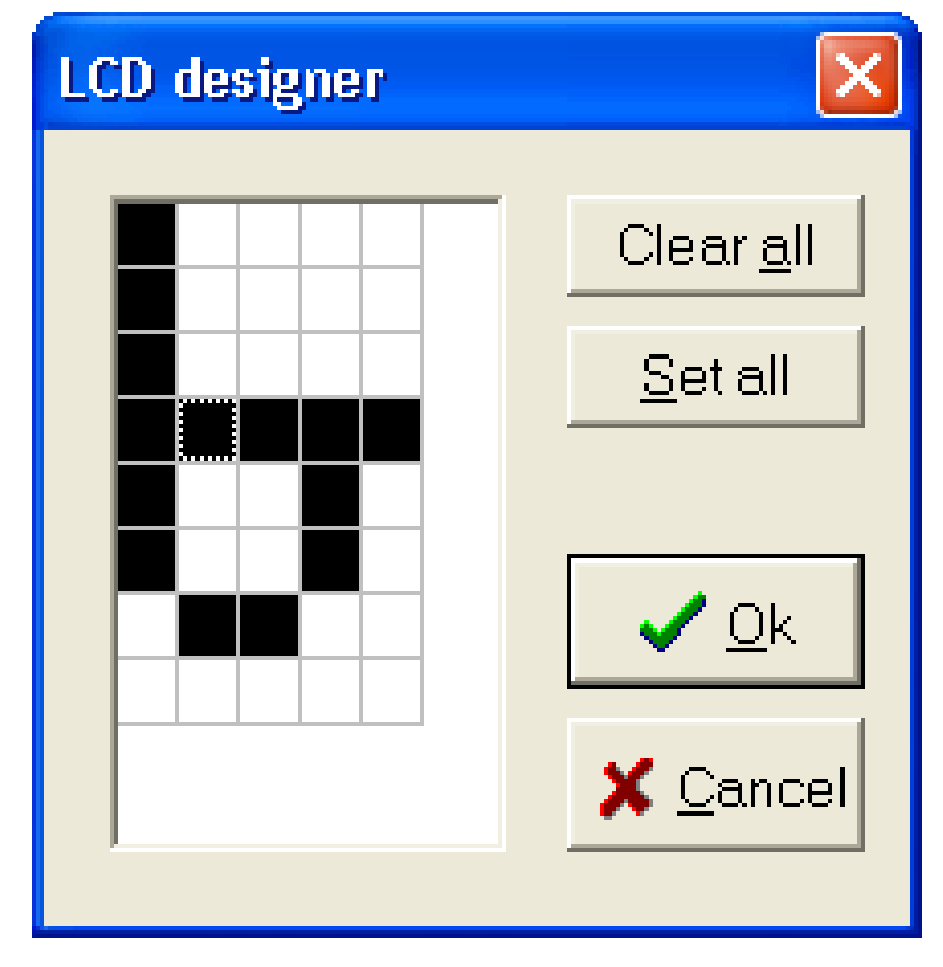}
	\end{center}
	\caption{ Bangla digit "8" is designed in the LCD designer  }
	\label{LCD designer2} 
\end{figure}

\begin{figure}[htbp]
	\begin{center}
	 \includegraphics[angle=0,scale=.4]{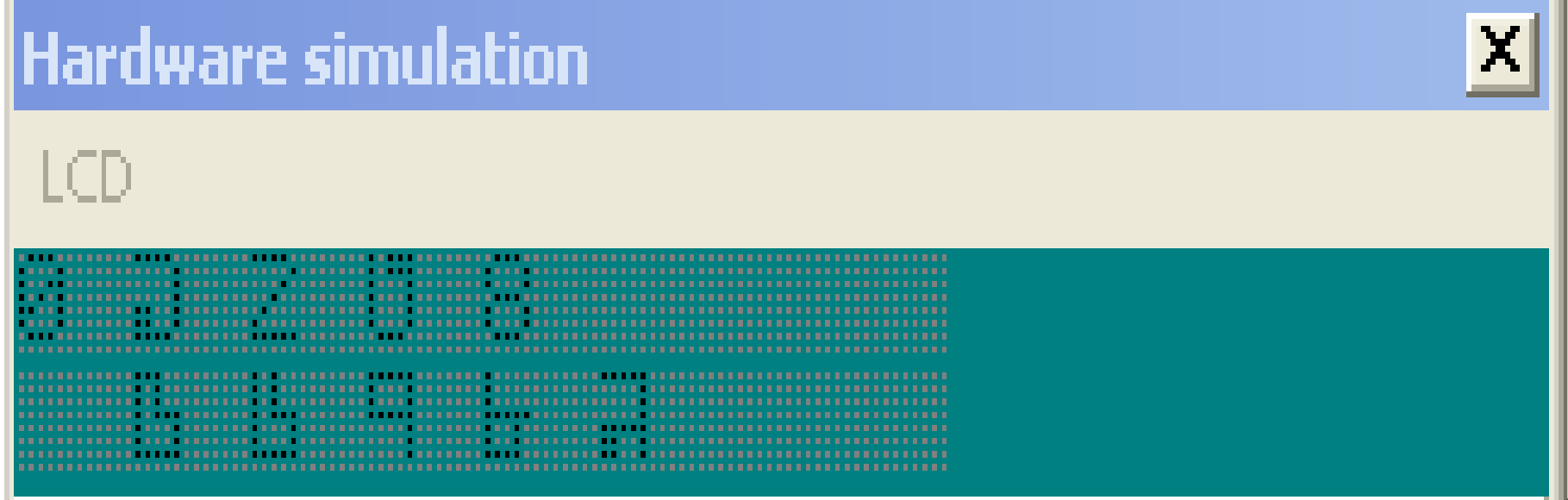}
	\end{center}
	\caption{LCD simulator - Ten Bangla digits }
	\label{LCD designer3} 
\end{figure}

\subsection{Clock algorithm design}
Clock algorithm design consists of two steps. 
	\begin{enumerate}
		\item Time counting process
		\item Time adjustment process
	\end{enumerate}
	
\subsubsection{Time counting process}	
 The responsibility of this process is to count time. At beginning of the process \textit{hour}, \textit{minute} and \textit{second} are set to zero. After passing each second, \textit{second} is incremented and checked if it exceeds fifty-nine. When \textit{second} exceeds fifty-nine then \textit{second} is set to zero and \textit{minute} is incremented. After passing one minute, it is checked if \textit{minute} exceeds fifty-nine. when \textit{minute} exceeds fifty-nine then \textit{minute} is set to zero and \textit{hour} is incremented. After passing one hour, it is checked if \textit{hour} exceeds twenty-three. When \textit{hour} exceeds twenty-three then \textit{hour} is set to zero and the process continues all over again. 
 
\subsubsection{Time adjustment process}
The responsibility of this process is to adjust time. This process starts when the user has pressed the \textit{SET} button. Then, it is checked if the \textit{INCREMENT} button is pressed. If  \textit{Yes} then \textit{hour} is incremented and checked if the value is twenty-four. If  \textit{Yes} then \textit{hour} is set to zero. On the other hand, if the \textit{DECREMENT} button is pressed then \textit{hour} is decremented and checked if the value is minus one. If  \textit{Yes} then \textit{hour} is set to twenty-three.\\

 The user again presses the \textit{SET} button for modifying minutes. It is checked if the \textit{INCREMENT} button is pressed. If \textit{Yes} then \textit{minute} is incremented and checked if the value is sixty. If \textit{Yes} then \textit{minute} is set to zero. If the \textit{DECREMENT} button is pressed then \textit{minute} is decremented and checked if the value is minus one. If \textit{Yes} then \textit{minute} is set to fifty-nine.\\

Similarly, the user again presses the \textit{SET} button for altering seconds. It is checked if the \textit{INCREMENT} button is pressed. If \textit{Yes} then \textit{second} is incremented and checked if the value is sixty. If \textit{Yes} then \textit{second} is set to zero. if the \textit{DECREMENT} button is pressed then \textit{second} is decremented and checked if the value is minus one. If \textit{Yes} then \textit{second} is set to fifty-nine.\\

Now, the user again presses the \textit{SET} button to stop the process.

\begin{figure*}[htbp]
	\begin{center}
	 \includegraphics[angle=0,scale=.8]{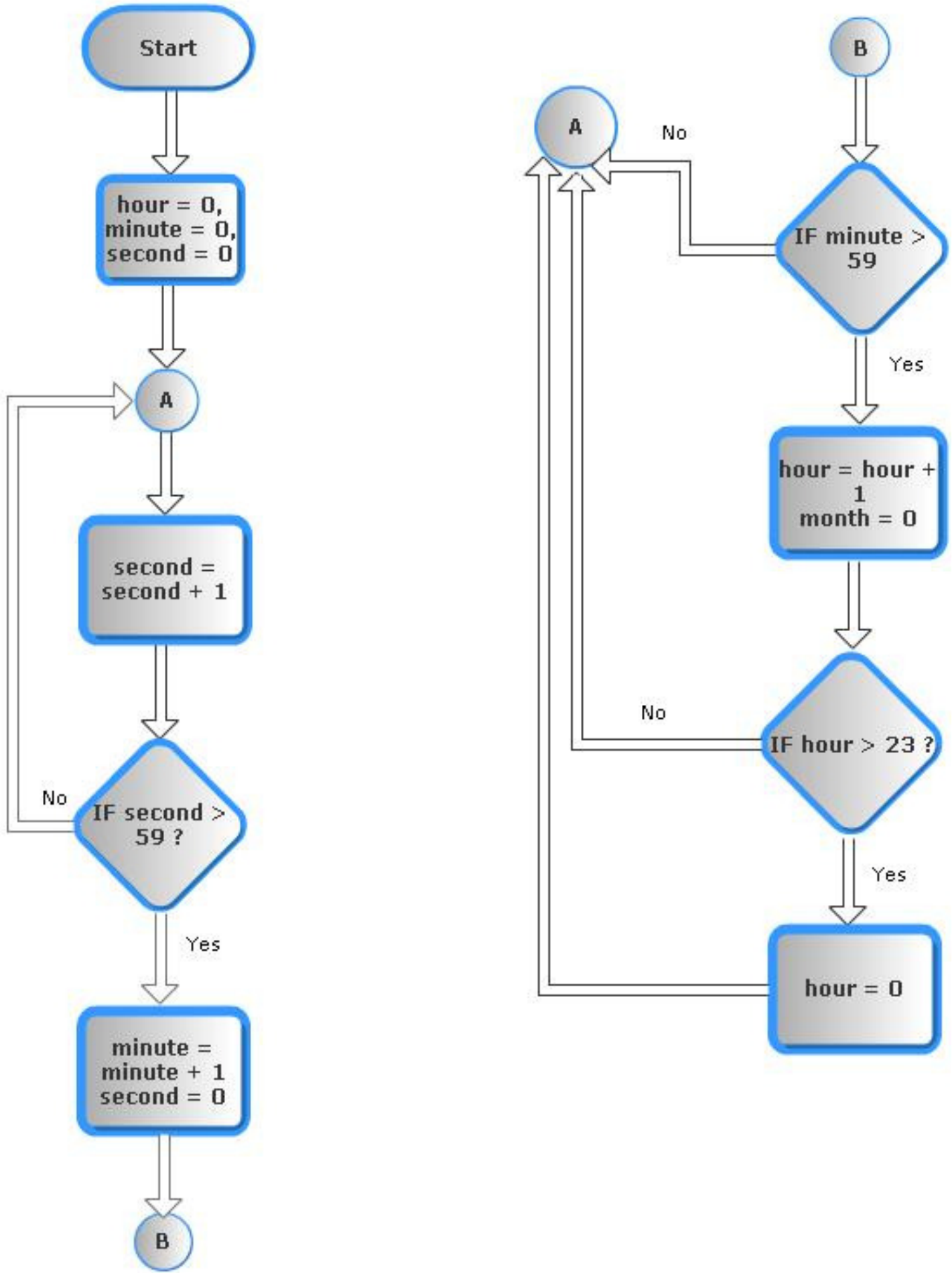}
	\end{center}
	\caption{ Time counting process flowchart }
	\label{timecountingprocess} 
\end{figure*}

\begin{figure*}[htbp]
	\begin{center}
	 \includegraphics[angle=0,scale=1.6]{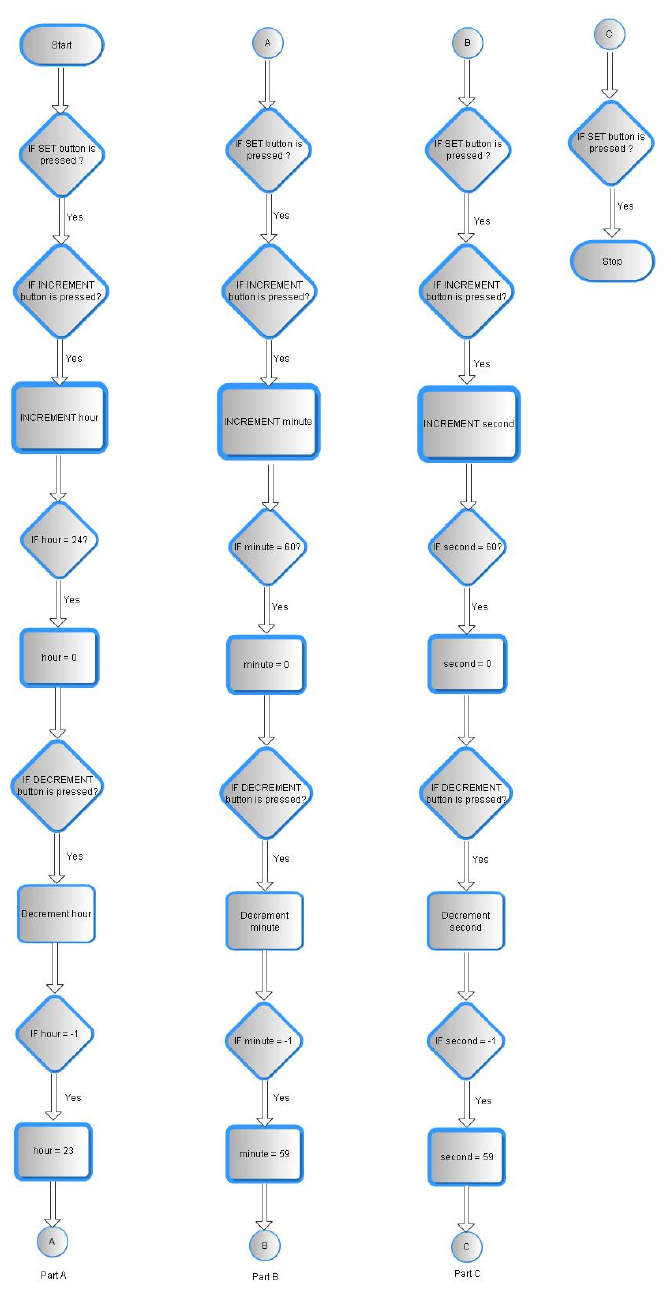}
	\end{center}
	\caption{ Time adjustment process flowchart  }
	\label{timeadjustmentprocess} 
\end{figure*}

\section{Design of digital clock}
The design of a clock circuit includes three units. They are input unit, processing unit and output unit. For input three push button switches are used which reset the time of the clock. For processing a microcontroller AT89C4051 \cite{Atmega128} is used. And for output 16*2 LCD display is used.

	\begin{figure*}[htbp]
	\begin{center}
	 \includegraphics[angle=0,scale=0.8]{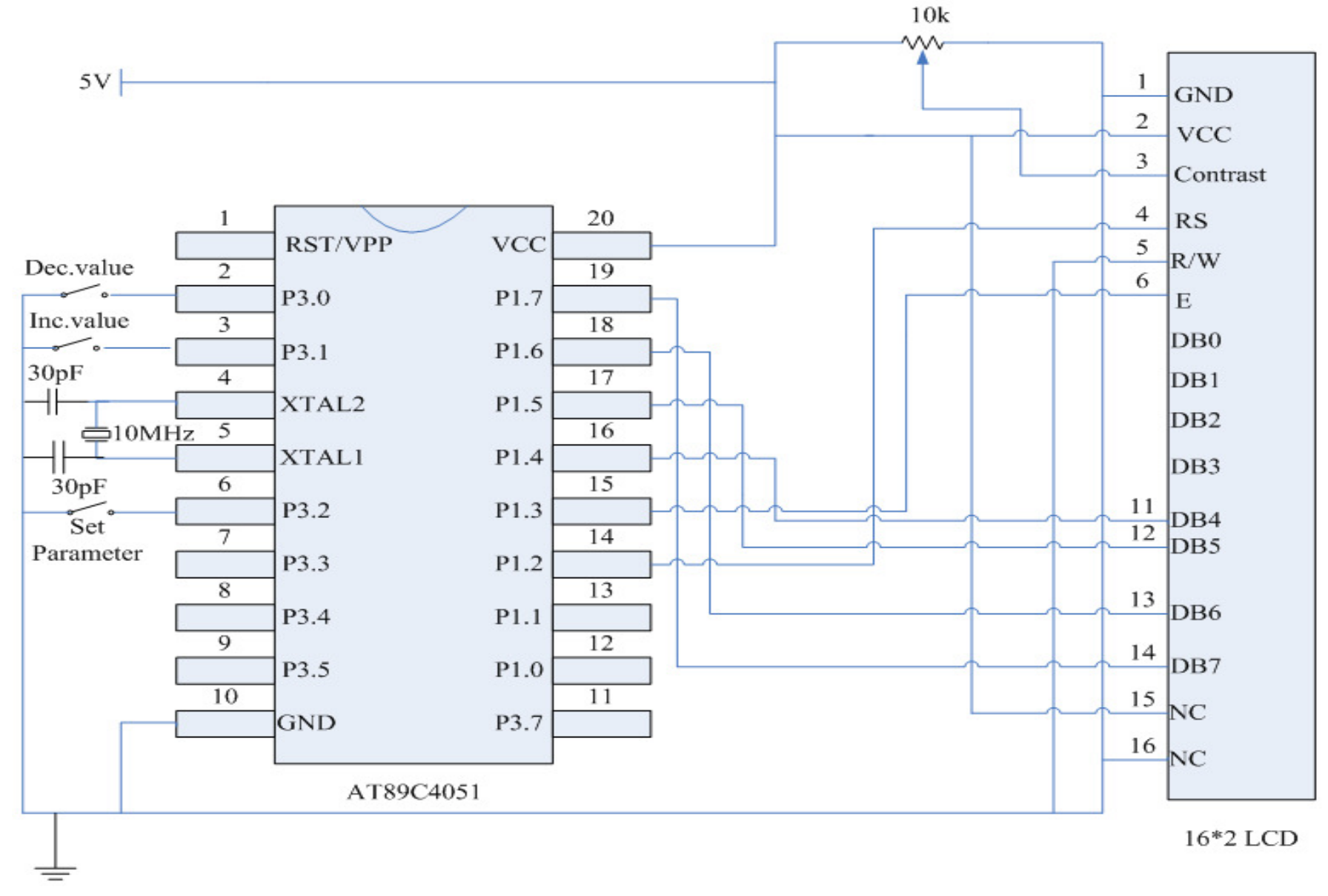}
	\end{center}
	\caption{ Circuit diagram of digital Bangla clock  }
	\label{digitalclockhardware} 
\end{figure*}

\subsection{List of circuit components}
\begin{center}
    \begin{tabular}{ | l | l |}
    \hline
    Component name & Quantity \\ \hline
    AT89C4051 & 1  \\ \hline
    16*2 LCD & 1  \\ \hline
    IN4001 diode & 1  \\ \hline
    11.06Mhz crystal oscillator & 1  \\ \hline
    10K potentiometer & 1  \\ \hline
    10K resistor & 3  \\ \hline
    8.2 resistor & 1  \\ \hline
    Push button switch & 3  \\ \hline
	30pF capacitor & 2  \\ \hline
	10uF capacitor & 2  \\ \hline
	0.1uF capacitor & 1  \\ \hline
	9V battery & 1  \\ \hline
    \end{tabular}
\end{center}

\subsection{Circuit description}
The three switches set, increment and decrement of the clock are connected to microcontroller through port P3.2, P3.1 and P3.0 shown in figure \ref{digitalclockhardware}. The crystal oscillator of the clock is connected to the microcontroller through pin no. 5 and 4. The upper four bits of the LCD  display is connected to microcontroller pin no. P1.7, P1.6, P1.5, P1.4. The enable, reset pin of the LCD is connected to the microcontroller with pin no. 15 and 14. The R/W pin of the microcontroller is connected to the ground. The contrast of the LCD is connected to the potentiometer. The power supply connection is connected to the pin no. 20 and ground connection is to the pin no. 10.

\section{Source code implementation}
The source code is written in BASIC language.
\subsection{Change hour}
\lstset{%
    backgroundcolor=\color{gray!50},
    basicstyle=\small\ttfamily\color{blue},%
    numbers=left, numberstyle=\tiny, stepnumber=1, numbersep=5pt,%
    tabsize=2 }%

\begin{lstlisting}[frame=single, framerule=0pt]
If P3.2 = 0 Then
 Do
  If P3.1 = 0 Then
   Incr Hh
    If Hh = 24 Then
     Hh = 0        	
    End If
  End If
			
  If P3.0 = 0 Then
   Decr Hh
   If Hh = -1 Then
    Hh = 23
   End If
  End If
			
  If P3.2 = 0 Then
   Exit Loop
  End If
 Loop
End If	

\end{lstlisting}

\subsection{Change minute}
\lstset{%
    backgroundcolor=\color{gray!50},
    basicstyle=\small\ttfamily\color{blue},%
    numbers=left, numberstyle=\tiny, stepnumber=1, numbersep=5pt,%
    }%

\begin{lstlisting}[frame=single, framerule=0pt]
If P3.2 = 0 Then
 Do
  If P3.1 = 0 Then
   Incr Mm
    If Mm = 60 Then
     Mm = 0
     End If
   End If
			
  If P3.0 = 0 Then
   Decr Mm
    If Mm = -1 Then
     Mm = 59
    End If
   End If
			
   If P3.2 = 0 Then
    Exit Loop
   End If
 Loop
End If	

\end{lstlisting}

\subsection{Change second}
\lstset{%
    backgroundcolor=\color{gray!50},
    basicstyle=\small\ttfamily\color{blue},%
    numbers=left, numberstyle=\tiny, stepnumber=1, numbersep=5pt,%
    }%

\begin{lstlisting}[frame=single, framerule=0pt]
If P3.2 = 0 Then
 Do
  If P3.1 = 0 Then
   Incr Ss
    If Ss = 60 Then
     Ss = 0
     End If
   End If
			
  If P3.0 = 0 Then
   Decr Ss
    If Ss = -1 Then
     Ss = 59
    End If
   End If
			
   If P3.2 = 0 Then
    Exit Loop
   End If
 Loop
End If	

\end{lstlisting}

\begin{figure}[htbp]
	\begin{center}
	 \includegraphics[angle=0,scale=.8]{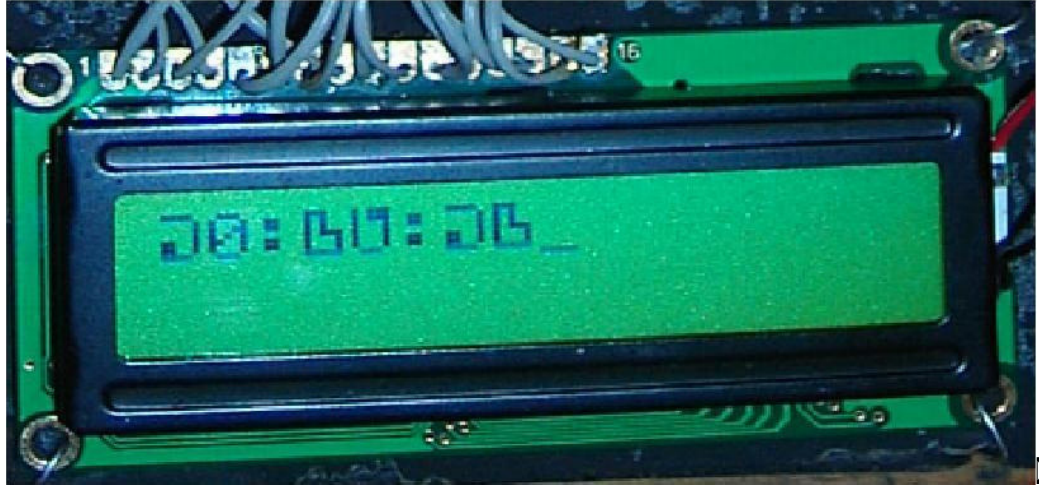}
	\end{center}
	\caption{ Digital clock showing digits in Bangla}
	\label{timecountingprocess} 
\end{figure}

\section{Summary and future work}

To implement a simple timer or counter is trivial, but to design and implement a digital clock showing real time is a complex task. The task becomes more complex when the font of the clock is Bangla. The design of the Bangla digtal clock is small and easy to use like other digital clocks. \\

It is possible to enhance the capability of this digital clock. With some hardware and software modification, it can become a fully automatic digital alarm clock showing digits in Bangla. This digital Bangla clock can be more useful if it can provide information like 12 hour clock or 24 hour clock, days of a week, month names, year etc. This features can be added very easily by software modification. In this design, 16*2 LCD display is used and first eight blocks is used to display second, minute and hour. So there are another eight blocks left to show additional information. The microcontroller AT89C4051 which has 4K memory, has enough memory space available for the additional programming of the microprocessor. The display of the digital Bangla clock can be more attractive by using graphic LCD.

\bibliography{lit}{}

\begin{thebibliography}{1}

\bibitem{lcd}
Vishay electronic gmbh, \textsf{{16 x 2 Character LCD}} datasheet.

\bibitem{Atmega128}
Atmel corporation, \textsf{{Atmel AT89C4051}} datasheet.

\bibitem{bascom}
Mcs electronics, \textsf{{BASCOM-AVR}}.

\bibitem{dallas}
Dallas semiconductor, \textsf{{DS12885 Real Time Clock}} datasheet.

\bibitem{avago}
Avago technologies, \textsf{{HDSP Series}} datasheet.

\end{thebibliography}
\bibliographystyle{plain}

\end{document}